\documentclass{ws-procs9x6-cpt16}
\usepackage{epstopdf}
\begin{document}

\newcommand{\refeq}[1]{(\ref{#1})}
\def\etal {{\it et al.}}

\title{Extra Dimensions and Violations of Lorentz Symmetry}

\author{James M.\ Overduin and Hamna Ali}

\address{Department of Physics, Astronomy and Geophysics, 
Towson University\\
Towson, MD 21252, USA}

\begin{abstract}
We use experimental limits on Lorentz violation to obtain new constraints on Kaluza-Klein-type theories in which the extra dimensions may be large but do not necessarily have units of length. The associated variation in fundamental quantities such as rest mass must occur slowly, on cosmological scales.
\end{abstract}

\bodymatter

\phantom{}\vskip10pt\noindent
Current approaches to unification of fundamental interactions based on extra dimensions generally assume that those dimensions are compact (as in string theories) or mere mathematical contrivances (as in projective theories), or that they are large but ``off limits'' to Standard-Model fields (as in brane theories). Here we explore the alternative idea that extra dimensions may be large but may not share the lengthlike character of the three macroscopic spatial dimensions (as in Space-Time-Matter theory).\cite{KKrefs}

\begin{figure}
\begin{center}
\includegraphics[width=\textwidth]{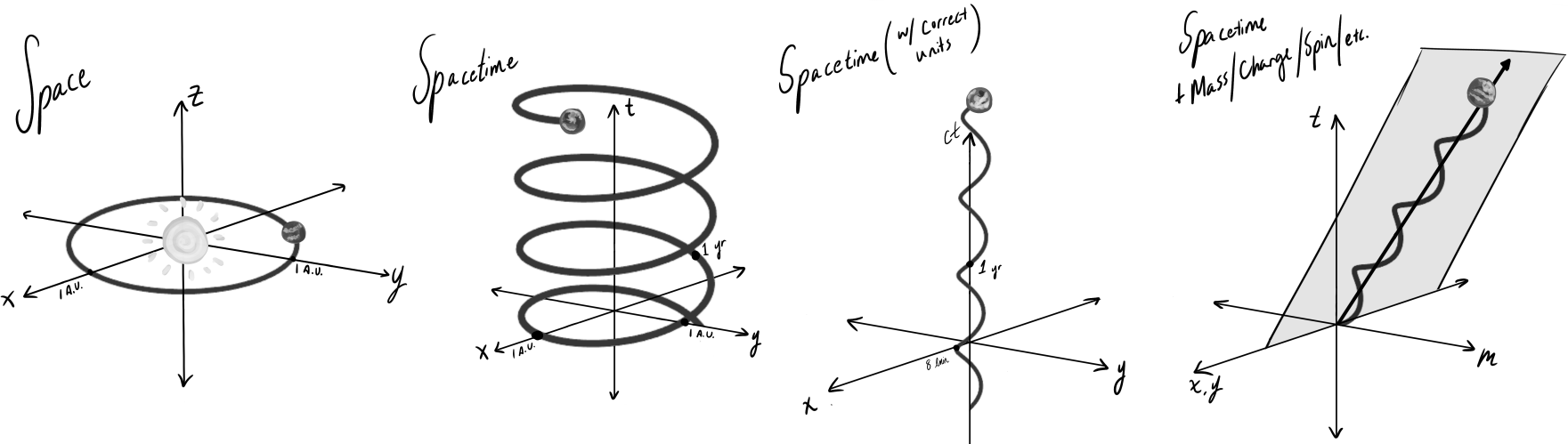}
\end{center}
\caption{Perspectives on extra dimensions. Far left, the Earth's orbit around the Sun in 3D space. Its trajectory in 4D spacetime is shown center left. In consistent units (center right), where displacement along the $t$-direction is measured in light years, the orbital radius is only eight light minutes and this trajectory is almost perfectly straight --- a geodesic in very nearly flat spacetime. Far right, a possible fourth dimension proportional to mass. ``Motion'' in this direction might manifest itself as a slow change in the rest masses of elementary particles (angle with respect to the $t$-axis greatly exaggerated).}
\label{fig1}
\end{figure}

Any fifth coordinate $x_4$ will introduce an additional term in the Lorentz factor of special relativity, as follows:
\begin{equation}
\gamma(v)=\left({1-\frac{v^2}{c^2}}\right)^{-1/2}\Rightarrow\left({1-\frac{v^2}{c^2}\pm\frac{v^2_4}{c^2}}\right)^{-1/2} \; ,
\label{gammaDefn}
\end{equation}
where $v_4 = dx_4/dt$ and we maintain an open mind with regard to signature. Consider the idea that $x_4$ might be proportional to rest mass $m$ (Fig.~\ref{fig1}).\cite{Wesson1984}
On dimensional grounds, $v_4=G\dot{m}/c^2$ where $\dot{m}\equiv dm/dt$. Experiment tells us that the new term must be small, 
so we can Taylor expand:
\begin{equation}
\gamma(v)=1+\frac{v^2}{2 c^2}\left[1\mp\left(\frac{G\dot{m}}{c^2v}\right)^2\right] \; .
\label{TaylorEq}
\end{equation}
Lorentz-violating terms of this kind occur in a comprehensive dynamical generalization of all known interactions termed the Standard-Model Extension or SME.\cite{ColladayKostelecky} It has been shown\cite{KosteleckyMewes} that the SME fully incorporates an earlier kinematical generalization of Special Relativity (SR) known as Robertson-Mansouri-Sexl (RMS) theory.\cite{RMSpapers} In RMS theory,  a preferred frame modifies the standard Lorentz transforms such that
$t=a(v)T+e(v)x,x=b(v)(X-vT),y=d(v)Y,z=d(v)Z$,
where $T,X,Y,Z$ are coordinates in the preferred frame, and the functions $a(v),b(v),d(v),e(v)$ describe time dilation, length contraction, transverse length contraction and clock synchronization respectively. $b(v)$ is a generalization of the Lorentz factor $\gamma(v)$ in Eq.~(\ref{gammaDefn}). Mansouri and Sexl showed on general grounds that
\begin{equation}
a(v)\sim 1+\alpha\frac{v^2}{c^2} \; , \;\;\; b(v)\sim 1+\beta\frac{v^2}{c^2} \; ,
\label{RMSeq}
\end{equation}
where $\alpha,\beta$ are constants whose values go over to $-\tfrac{1}{2},+\tfrac{1}{2}$ in the SR limit and 0,0 in the limit of Galilean relativity. 
Constraints on the RMS parameters $\alpha,\beta$ come from tests of the relativistic Doppler effect, known as the Ives-Stilwell (IS) experiment; and from the Kennedy-Thorndike (KT) experiment, a modified form of the original Michelson-Morley experiment with arms of different length. Recent limits are
$|\alpha+\tfrac{1}{2}|\leqslant8.4\times10^{-8}$ (IS)\cite{TobarEtal} and $\alpha-\beta+1=0.0^{+3.7}_{-4.8}\times10^{-8}$ (KT).\cite{ReinhardtEtal}
Combining these expressions, and comparing Eq.~(\ref{TaylorEq}) for $\gamma(v)$ with Eq.~(\ref{RMSeq}) for $b(v)$, we arrive at
\begin{equation}
|\beta-\tfrac{1}{2}|=\frac{1}{2}\left(\frac{G\dot{m}}{c^2v}\right)^2\leqslant1\times10^{-7} \; ,
\label{betaConstraint}
\end{equation}
from which it follows that $|\dot{m}|\leqslant2\times10^{32}$~kg/s (assuming that $v\leqslant c$). The weakness of this constraint follows from the tiny value of the dimension-transposing constant $G/c^3$ in everyday units.

A possible interpretation of the constraint~(\ref{betaConstraint}) is that there is a slow variation in the masses of elementary particles on cosmological timescales. Indeed, dividing by the mass of the observable universe, $M\sim\frac{4}{3}\pi\rho_{\mbox{\tiny crit}}(ct_0)^3=c^3t_0/2G$ (where $\rho_{\mbox{\tiny crit}}=3H_0^2/8\pi G$, $H_0$ is the Hubble expansion rate, $t_0$ the age of the Universe, and $H_0t_0\approx1$ from observation), we find that
$\dot{m}/M\lesssim{H_0}/{1000}$.
On this interpretation, the origin of mass might be attributed dynamically to a fifth dimension. This could perhaps be regarded as a geometrical counterpart to the Higgs mechanism, with the degree of freedom inherent in a scalar field being associated instead with a new coordinate. Related ideas have been explored by others.\cite{relatedIdeas}

There is a satisfying symmetry in placing mass on the same footing as space and time, since these are the three base dimensions of fundamental physics. Other possibilities exist as well. The fundamental constants  give us three possible dimension-transposing factors $G,c,\hbar$, which we might guess involve gravity, relativity and particle physics respectively. (Alternatively, it is suggestive that black holes are characterized by precisely three properties: mass, charge and spin.) We have carried out analogous calculations for extra dimensions related to mass $m$ by $x_4\propto\hbar/mc$ to charge $q$ via $x_5\propto\sqrt{\hbar G/c^3}(q/e)$
and to spin $\ell$ by $x_6\propto\sqrt{G/\hbar c^3}\ell$,
with similar conclusions in each case.
In future work, we hope to report more fully on these possibilities, and to express our results in terms of SME parameters.\cite{KosteleckyMewes2009}

\section*{Acknowledgments}
H.A. thanks the Department of Physics, Astronomy and Geosciences and Fisher College of Science and Math, Towson University, for travel support.

\end{document}